\documentclass[aps, prd, superscriptaddress, preprintnumbers, floatfix, longbibliography,10pt,twocolumn]{revtex4-2}

\usepackage{amsfonts,amsmath,amssymb,amstext}
\usepackage{graphicx,epstopdf}
\usepackage{float,wrapfig}
\usepackage{subfigure, psfrag}
\usepackage{dsfont,bm}
\usepackage{color}
\usepackage[normalem]{ulem}
\usepackage[colorlinks=true,urlcolor=blue,citecolor=blue,linkcolor=blue]{hyperref}
\hypersetup{breaklinks=true}
\usepackage{verbatim}

\usepackage{array}
\newcolumntype{L}[1]{>{\raggedright\let\newline\\\arraybackslash\hspace{0pt}}m{#1}}
\newcolumntype{C}[1]{>{\centering\let\newline\\\arraybackslash\hspace{0pt}}m{#1}}
\newcolumntype{R}[1]{>{\raggedleft\let\newline\\\arraybackslash\hspace{0pt}}m{#1}}

\definecolor{purple}{rgb}{0.8,0,0.6}
\definecolor{darkgreen}{rgb}{0.00,0.6,0.00}

\usepackage[usenames,dvipsnames]{xcolor}

\extrafloats{100}

\begin{document}

\title{Stackings and effective models of bilayer dice lattices}
\date{August 31, 2023}

\author{P.~O.~Sukhachov}
\email{pavlo.sukhachov@yale.edu}
\affiliation{Department of Physics, Yale University, New Haven, Connecticut 06520, USA}

\author{D.~O.~Oriekhov}
\affiliation{Instituut-Lorentz, Universiteit Leiden, P.O. Box 9506, 2300 RA Leiden, The Netherlands}

\author{E.~V.~Gorbar}
\affiliation{Faculty of Physics, Kyiv National Taras Shevchenko University, 64/13 Volodymyrska st., 01601 Kyiv, Ukraine}
\affiliation{Bogolyubov Institute for Theoretical Physics, 14-b Metrolohichna st., 03143 Kyiv, Ukraine}

\begin{abstract}
We introduce and classify nonequivalent commensurate stackings for bilayer dice or $\mathcal{T}_3$ lattice. For each of the four stackings with vertical alignment of sites in two layers, a tight-binding model and an effective model describing the properties in the vicinity of the threefold band-crossing points are derived. Focusing on these band-crossing points, we found that although the energy spectrum remains always gapless, depending on the stacking, different types of quasiparticle spectra arise. They include those with flat, tilted, anisotropic semi-Dirac, and $C_3$-corrugated energy bands. We use the derived tight-binding models to calculate the density of states and the spectral function. The corresponding results reveal drastic redistribution of the spectral weight due to the inter-layer coupling that is unique for each of the stackings.
\end{abstract}

\maketitle

\section{Introduction}
\label{sec:Introduction}

The search for novel materials with unusual dispersion relations is one of the major topics in modern condensed matter physics. There are several successful examples of this search that lead to vigorous research directions. Among them, graphene is, perhaps, the most well-known example of a solid with an unusual dispersion relation. Indeed, at low energies, graphene's electron quasiparticles are described by a two-dimensional (2D) Dirac equation~\cite{Novoselov:2004,Geim-Novoselov:rev-2007,Katsnelson:book-2012}. The 2D Dirac spectrum can be also realized at the surface of three-dimensional (3D) topological insulators~\cite{Hsieh-Hasan:2008,Hasan-Kane:rev-2010,Bernevig:book-2013}. Finally, the 3D linear energy spectrum appears in Weyl and Dirac semimetals~\cite{Kim-Li-BiSb:2013,Huang-Chen-TaAs:2015,Yang-Zu-NbAs:2015,Armitage:rev-2018,GMSS:book}.

Intermediate between 2D and 3D materials are layered systems. The energy spectrum of these systems can be engineered by stacking the layers in a certain order. The electronic properties of the corresponding few-layer systems can be drastically different from their single-layer counterparts. For example, bilayer graphene in the Bernal ($A-B$) stacking reveals a quadratic quasiparticle spectrum in the vicinity of band touching points~\cite{Novoselov-Geim:2006,McCann-Koshino:2012,Katsnelson:book-2012}. This leads to a different integer quantum Hall effect~\cite{Novoselov-Geim:2006,McCann-Falko:2006} and optical response~\cite{Abergel-Falko:2006} compared to single-layer graphene.

Recently, there has been a surge of interest in materials containing even more exotic energy spectra with flat bands. Among these systems, perhaps the most well known is twisted bilayer graphene (TBG)~\cite{LopesdosSantos-CastroNeto:2007,Morell-Barticevic:2010,Bistritzer-MacDonald:2010,Cao-Jarillo-Herrero:2018-TBG,Yankowitz-Dean:2018,Lu-Efetov:2019,Cao-Jarillo-Herrero:2020}; see also Ref.~\cite{Andrei-MacDonald:2020-rev} for a review. In essence, TBG is composed of two layers of graphene rotated with respect to each other by some angle. It was shown~\cite{Morell-Barticevic:2010,Bistritzer-MacDonald:2010} that for the specific, so-called ``magic" twist angles, 2D isolated flat bands appear in the energy spectrum of TBG. The presence of flat bands is directly related to the nontrivial properties of TBG including interaction effects such as superconductivity near integer band-filling factors~\cite{Cao-Jarillo-Herrero:2018-TBG,Yankowitz-Dean:2018,Lu-Efetov:2019,Cao-Jarillo-Herrero:2020,Oh-Yazdani:2021}.

While TBG receives significant attention nowadays, historically, the appearance of flat bands was predicted a few decades ago in kagome~\cite{Syozi:1951}, dice or $\mathcal{T}_3$~\cite{Sutherland:1986,Vidal-Doucot:1998}, and Lieb~\cite{Lieb:1989} lattices. A kagome lattice consists of equivalent lattice points and equivalent bonds forming equilateral triangles and regular hexagons; each hexagon is surrounded by triangles and vice versa. A Lieb lattice is described by three sites in a square unit cell where two of the sites are neighbored by two other sites and the third site has four neighbors. In essence, a dice lattice has a hexagonal structure with an additional site placed in the center of each hexagon. The central site acts as a hub connected to six rims while each of the rims is connected to three hubs; see also Fig.~\ref{fig:Model-dice}(a) for a dice lattice. If one of the rims is removed, a conventional honeycomb (graphene) lattice is restored. In the rest of this work, we focus on a dice lattice as a representative system. As for experimental setups, dice lattices were proposed in artificial systems such as optical lattices~\cite{Rizzi-Fazio:2005,Bercioux-Haeusler:2009}; see Ref.~\cite{Leykam-Flach:2018} for a review. As an example of the experimental realizations of dice lattices, we mention Josephson arrays~\cite{Serret-Pannetier:2002} as well as optical realizations~\cite{Rizzi-Fazio:2005}.

The lattice structure of the dice model with three sites per unit cell leads to three bands in the energy spectrum which is similar to that in graphene albeit with Dirac points intersected by a flat band~\cite{Raoux-Montambaux:2013}. The corresponding low-energy spectrum can be described in terms of spin-1 fermions, which have no analogs in high-energy physics. The flat band leads to strikingly different physical properties with a paramagnetic response~\cite{Raoux-Montambaux:2013,Piechon-Montambaux:2015} instead of the diamagnetic one as in graphene~\cite{McClure:1956} being a representative example. To the best of our knowledge, multi-layer dice lattices were not investigated before and, as in multi-layer graphene, are expected to be different from their single-layer counterparts.

In this work, we combine two vigorous research directions related to exotic lattices and heterostructures by studying the properties of \emph{bilayer dice lattices}~\footnote{Bilayer dice lattices should not be confused with the double-layer lattice studied in Ref.~\cite{Iurov-Huang:2020}.}. We classify nonequivalent commensurate stackings of dice lattices with aligned sites and formulate the corresponding tight-binding and effective models. The latter describe the properties of the threefold band-crossing points. Depending on the type of the stacking, the spectrum in the vicinity of these band-crossing points comprises Dirac points intersected by flat bands, $C_3$-corrugated bands, tilted bands, or even a semi-Dirac spectrum. For the semi-Dirac spectrum, the energy bands are anisotropic with a linear dispersion relation along one direction and the quadratic dispersion along the other~\cite{Hasegawa-Kohmoto:2006}. For all four nonequivalent stackings, the sets of band-crossing points originating from different layers are separated in energy with the separation determined by the interlayer coupling constant. The obtained bilayer models are illustrated by calculating the density of states (DOS) and the spectral function. Being strongly modified by the interlayer coupling, the DOS and the spectral function provide an efficient way to distinguish the stackings and set the stage for the investigation of the optical response in our work~\cite{SOG:part2-2023}.

The paper is organized as follows. We discuss the key properties of a single-layer dice lattice in Sec.~\ref{sec:SL}. The commensurate stackings are classified, and the tight-binding and effective models of a bilayer dice lattice are formulated in Sec.~\ref{sec:BL}. The spectral functions and the DOS for each of the four stackings are presented in Sec.~\ref{sec:DOS}. The results are summarized in Sec.~\ref{sec:Summary}. Technical details concerning the derivation of the effective models, spectral functions, and the properties of the bilayer lattices at larger coupling constants are presented in Appendices~\ref{sec:App-Model} and \ref{sec:App-spectral-low}, respectively.

\section{Single-layer dice lattice}
\label{sec:SL}

\begin{figure*}[t]
\centering
\subfigure[]{\includegraphics[width=0.3\textwidth]{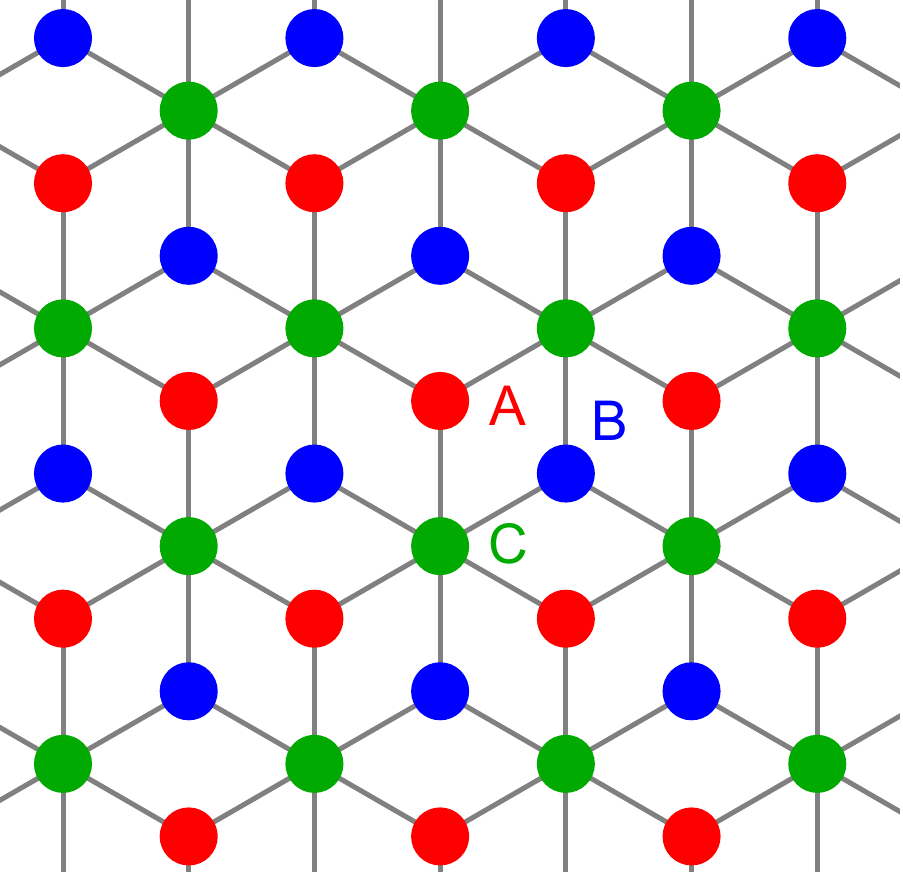}}
\hspace{0.1\textwidth}
\subfigure[]{\includegraphics[width=0.45\textwidth]{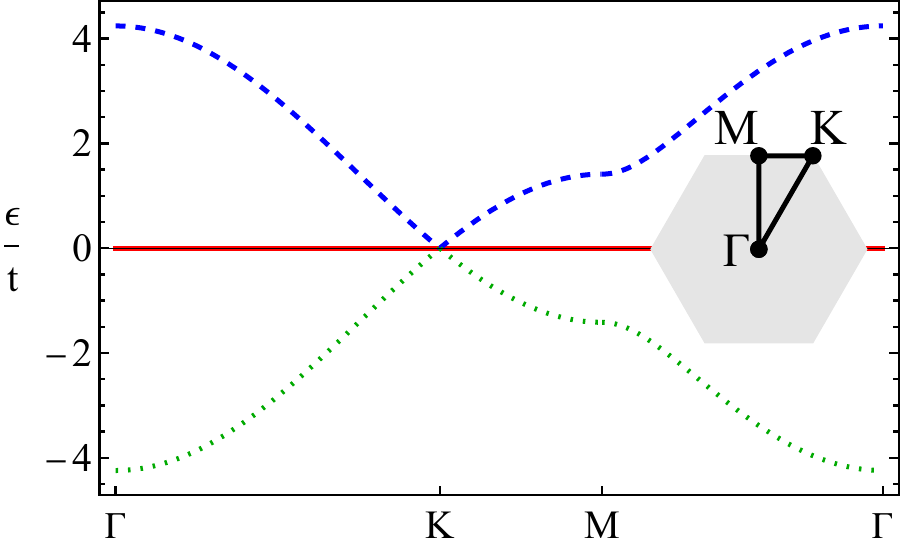}}
\caption{
Panel (a): The schematic representation of single-layer dice lattice. The $A$, $B$, and $C$ sites are denoted by red, blue, and green dots.
Panel (b): The energy spectrum given in Eq.~(\ref{Model-tb-energy-0-pm}) along the $\Gamma-\mbox{K}-\mbox{M}-\Gamma$ line in the Brillouin zone (inset). Here, $t$ is the hopping constant.
}
\label{fig:Model-dice}
\end{figure*}

As a warm-up and to set the stage for the discussion of the bilayer dice lattice, we present the model and the key properties of a \emph{single-layer dice lattice}. In essence, a dice lattice is a hexagonal lattice composed of two sublattices (denoted as $A$ and $B$) with additional sites ($C$ sublattice) placed in the center of hexagons. The resulting inter-sublattice connections are shown in Fig.~\ref{fig:Model-dice}(a). As one can see, the atoms of the $C$ sublattice act as hubs that connect to six neighbors, while the atoms of the $A$ and $B$ sublattices (rims) connect only to three neighbors.

In the basis of states corresponding to the $A$, $C$, and $B$ sublattices, the tight-binding Hamiltonian reads~\cite{Raoux-Montambaux:2013}
\begin{equation}
\label{Model-tb-h-def}
H(\mathbf{q}) = \left(
                  \begin{array}{ccc}
                    0 & -t \sum_{j}e^{-i\mathbf{q}\cdot\bm{\delta}_j} & 0 \\
                    -t \sum_{j}e^{i\mathbf{q}\cdot\bm{\delta}_j} & 0 & -t \sum_{j}e^{-i\mathbf{q}\cdot\bm{\delta}_j} \\
                    0 & -t \sum_{j}e^{i\mathbf{q}\cdot\bm{\delta}_j} & 0 \\
                  \end{array}
                \right),
\end{equation}
where $t$ is the hopping constant, $\mathbf{q}$ is the wave vector in the Brillouin zone, and
\begin{equation}
\label{Model-tb-delta-1-2-3}
\bm{\delta}_1 = a \left\{0,1\right\}, \,\,
\bm{\delta}_2 = a \left\{\frac{\sqrt{3}}{2},-\frac{1}{2}\right\}, \,\,
\bm{\delta}_3 = a \left\{-\frac{\sqrt{3}}{2},-\frac{1}{2}\right\}
\end{equation}
denote the relative positions of sites $A$ with respect to sites $C$; $a$ is the distance between neighboring $A$ and $C$ sites. The same vectors but with the minus sign denote the relative positions of sites $B$ with respect to sites $C$. In this model, the $A$ and $B$ sublattices are equivalent.

The energy spectrum of Hamiltonian (\ref{Model-tb-h-def}) reads
\begin{widetext}
\begin{equation}
\label{Model-tb-energy-0-pm}
\epsilon_{0} = 0, \quad \quad \epsilon_{\pm} = \pm t \sqrt{6} \sqrt{1 + \frac{2}{3} \cos{\left(\sqrt{3} aq_x\right)} +\frac{4}{3} \cos{\left(\frac{\sqrt{3}}{2} aq_x\right)}\cos{\left(\frac{3}{2} aq_y\right)}}.
\end{equation}
\end{widetext}
The dispersive bands $\epsilon_{\pm}$ are the same as in graphene where the quasiparticle spectrum contains two nonequivalent Dirac points $K$ and $K^{\prime}$. We show the corresponding energy spectrum in Fig.~\ref{fig:Model-dice}(b).

In the vicinity of the Dirac points, Hamiltonian (\ref{Model-tb-h-def}) can be linearized and reads as
\begin{equation}
\label{Model-tb-h-lin}
H_{\xi}(\mathbf{k}) = \hbar v_F \left(\xi S_x k_x +S_y k_y\right),
\end{equation}
where $\mathbf{k} = \mathbf{q} -\mathbf{K}_{\xi}$ is the wave vector measured relative to the $K$ ($\xi=+$) and $K^{\prime}$ ($\xi=-$) points located at $\mathbf{K}_{\xi} = \xi 4\pi /(3\sqrt{3}a) \left\{1,0\right\}$ and $v_F = 3ta/(\sqrt{2}\hbar)$ is the Fermi velocity. Further, we introduced the following spin-1 matrices:
\begin{equation}
\label{Model-S-def}
S_x = \frac{1}{\sqrt{2}}\left(
        \begin{array}{ccc}
          0 & 1 & 0 \\
          1 & 0 & 1 \\
          0 & 1 & 0 \\
        \end{array}
      \right),  \quad
      S_y = \frac{1}{\sqrt{2}} \left(
        \begin{array}{ccc}
          0 & -i & 0 \\
          i & 0 & -i \\
          0 & i & 0 \\
        \end{array}
      \right).
\end{equation}
The corresponding energy spectrum contains a Dirac point intersected by a flat band
\begin{equation}
\label{Mode-energy-lin}
\epsilon_{0} = 0, \quad \quad \quad \epsilon_{\pm} = \pm \hbar v_F k.
\end{equation}

As we discussed in Sec.~\ref{sec:Introduction}, heterostructures made of different stackings of single-layer graphene are a major topic in graphene physics. In the next section, we will introduce and study the simplest multilayer dice lattices composed of two commensurately stacked single-layer dice lattices.

\section{Bilayer dice lattice}
\label{sec:BL}

\subsection{Stackings of bilayer dice lattices}
\label{sec:BL-stackings}

For the \emph{bilayer dice lattice}, there are a few ways to commensurately stack two dice lattices with vertically aligned sites. The most obvious way is to have the sublattices of the same type in two layers aligned with each other. Therefore, we call this type of stacking the \emph{aligned $AA-BB-CC$ stacking}. Other stackings can be obtained starting from the aligned stacking by rotating or shifting one of the layers. A commensurate stacking is obtained by rotating one of the layers around a $C$ site by $\pi/3$. In this case, the sublattices $A$ and $B$ in one of the layers are aligned with the sublattices $B$ and $A$ of the other layer. Because the hub atoms $C$ remain aligned, we dub this type of the stacking the \emph{hub-aligned $AB-BA-CC$ stacking}. We notice that the $A$ and $B$ sublattices have different connectivity compared to the $C$ sublattice. Therefore, a nonequivalent stacking is realized for rotating around an $A$ site by $\pi/3$; rotation around a $B$ site (with the resulting $AC-BB-CA$ stacking) is equivalent since the sublattices $A$ and $B$ are assumed to be interchangeable within each of the layers. This results in the \emph{mixed $AA-BC-CB$ stacking} where the sublattices $B$ and $C$ in one layer are aligned with the sublattices $C$ and $B$ in the other; i.e., hubs and rims intermix. Finally, we can shift one of the layers with respect to the other by the distance between neighboring $A$ and $C$ sites. For the corresponding commensurate stacking, the sublattices $A$, $B$, and $C$ in one layer are aligned with the sublattices $C$, $A$, and $B$ in the other. We call this type of stacking the \emph{cyclic $AB-BC-CA$ stacking}.
Other stackings are either equivalent, noncommensurate, or have misaligned sites.

Certainly, it would be interesting to determine which of these stackings has the lowest energy. Unlike bilayer graphene, where the Bernal $A-B$ stacking is more energetically favorable than the $A-A$ one~\cite{Mostaani-Falko:2015}, lattice sites of the bilayer dice lattices considered in this work are always aligned with each other. This suggests that the stacking energy is not much different for these stackings. To address this question, however, a more refined analysis that depends on a particular realization of the dice lattice is required.

Thus, there are four nonequivalent vertically aligned commensurate stackings in a bilayer dice lattice:
(i) aligned $AA-BB-CC$, (ii) hub-aligned $AB-BA-CC$, (iii) mixed $AA-BC-CB$, and (iv) cyclic $AB-BC-CA$. We model interlayer hoppings in these stackings by the following interlayer-coupling Hamiltonians:
\begin{eqnarray}
\label{Model-h-tun-1-2-3-4}
H_{\rm c}^{\rm (a)} &=& g \left(
                  \begin{array}{ccc}
                    1 & 0 & 0 \\
                    0 & 1 & 0 \\
                    0 & 0 & 1 \\
                  \end{array}
                \right), \quad
                H_{\rm c}^{\rm (h)} = g \left(
                  \begin{array}{ccc}
                    0 & 0 & 1 \\
                    0 & 1 & 0 \\
                    1 & 0 & 0 \\
                  \end{array}
                \right), \nonumber \\
                H_{\rm c}^{\rm (m)} &=& g \left(
                  \begin{array}{ccc}
                    1 & 0 & 0 \\
                    0 & 0 & 1 \\
                    0 & 1 & 0 \\
                  \end{array}
                \right), \quad
                H_{\rm c}^{\rm (c)} = g \left(
                  \begin{array}{ccc}
                    0 & 0 & 1 \\
                    1 & 0 & 0 \\
                    0 & 1 & 0 \\
                  \end{array}
                \right),
\end{eqnarray}
where $g$ is the coupling constant. In writing Eq.~(\ref{Model-h-tun-1-2-3-4}), we assumed only the nearest-neighbor tunneling. For simplicity, the coupling constants for all sites are taken to be equivalent.

The tight-binding Hamiltonian for a bilayer dice lattice reads as
\begin{equation}
\label{Model-H-total}
H_{\rm tot}(\mathbf{q}) = \left(
                  \begin{array}{cc}
                    H(\mathbf{q}) & H_{\rm c} \\
                    H_{\rm c}^{\rm T} & H(\mathbf{q}) \\
                  \end{array}
                \right),
\end{equation}
where $H(\mathbf{q})$ is given by the single-layer tight-binding Hamiltonian (\ref{Model-tb-h-def}) and $H_{\rm c}$ is defined by one of the coupling Hamiltonians in Eq.~(\ref{Model-h-tun-1-2-3-4}).

Before discussing the effective models, it is instructive to analyze the discrete symmetries of the tight-binding Hamiltonian (\ref{Model-H-total}) and compare them with their counterparts in a single-layer dice lattice.

\subsection{Discrete symmetries}
\label{sec:BL-symmetries}

Discrete symmetries including charge-conjugation, time-reversal, and inversion symmetries play an important role in many condensed matter systems allowing for the classification of electron states and order parameters. The single-layer dice lattice respects all of these symmetries as well as possesses the $C_3$ rotational symmetry. The coupling Hamiltonian of the bilayer lattice might, however, break one or more of the discrete symmetries. We summarize the symmetries in Table~\ref{tab:symmetry} and provide a more detailed discussion below.

\begin{table*}[!ht]
	\begin{tabular}{|L{4.5cm}|C{4.5cm}|C{4.cm}|C{4.cm}|}
		\hline
		Dice lattice & Charge-conjugation symmetry  & Time-reversal symmetry & Inversion and in-plane inversion symmetries\\ \hline
		Single-layer & $M_0\hat{K}$ & $\mathds{1}_3\hat{K}$ & $W_0$ \\ \hline
		Aligned $AA-BB-CC$ & $M_1 \hat{K}$,\,\, $M_2 \hat{K}$ & $\mathds{1}_3\hat{K}$ & $W_2$, \,\, $W_1$ \\ \hline
		Hub-aligned $AB-BA-CC$ &  $M_1\hat{K}$,\,\, $M_2\hat{K}$ & $\mathds{1}_3\hat{K}$ & $W_2$, \,\, $W_1$ \\ \hline
		Mixed $AA-BC-CB$ & - & $\mathds{1}_3\hat{K}$ & - \\ \hline
		Cyclic $AB-BC-CA$ & - & $\mathds{1}_3\hat{K}$ & $W_2$ \\
		\hline
	\end{tabular}
\caption{Symmetry properties of the tight-binding Hamiltonian for a bilayer dice lattice (\ref{Model-H-total}) in different commensurate stackings. The tight-binding Hamiltonian of a single-layer dice lattice is given in Eq.~(\ref{Model-tb-h-def}) and the coupling Hamiltonians are defined in Eq.~(\ref{Model-h-tun-1-2-3-4}). The symmetry matrices $M_{1,2}$ and $W_{1,2}$ are defined in Eqs.~(\ref{eq:C-sym-M-matrix}) and (\ref{Model-symmetry-P-matrix}).
}
\label{tab:symmetry}
\end{table*}

We begin our symmetry analysis with the charge-conjugation or particle-hole symmetry ($\mathcal{C}$-symmetry). The operator of the charge-conjugation symmetry is defined as
\begin{equation}
\label{Model-symmetry-C}
\hat{C}\hat{H}(\mathbf{q})\hat{C}^{-1} = -\hat{H}(\mathbf{q}),
\end{equation}
where $\hat{H}(\mathbf{q})$ is the second-quantized version of the Hamiltonian $H(\mathbf{q})$.
The corresponding operator necessarily contains the complex conjugation operator $\hat{K}$ and a matrix, i.e., $\hat{C}= M\hat{K}$~\footnote{Notice that the complex conjugation operator $\hat{K}$ should be absent if we use the first-quantized version of the Hamiltonian in Eq.~(\ref{Model-symmetry-C}); see, e.g., Ref.~\cite{Zirnbauer:2021} for a detailed explanation of the particle-hole and charge-conjugation symmetries.}. For the aligned $AA-BB-CC$ and hub-aligned $AB-BA-CC$ stackings, there are the following matrices $M$:
\begin{equation}
\label{eq:C-sym-M-matrix}
M_1 = \tau_z\otimes M_0, \quad
M_2 = i\tau_y\otimes M_0, \quad
M_0 = \begin{pmatrix}
	0 & 0 & 1\\
	0 & -1& 0\\
	1 & 0 & 0
\end{pmatrix}.
\end{equation}
Here, $\bm{\tau}$ is the vector of the Pauli matrices defined in the layer space and $M_0$ is the charge-conjugation symmetry matrix for a single-layer dice lattice~\cite{Gorbar-Oriekhov:2021}. No charge-conjugation symmetry exists for the mixed $AA-BC-CB$ and cyclic $AB-BC-CA$ stackings.
	
Let us proceed to the time-reversal symmetry ($\mathcal{T}$ symmetry), which is defined as
\begin{equation}
\label{Model-symmetry-T}
\hat{T} \hat{H}(\mathbf{q}) \hat{T}^{-1} = \hat{H}(-\mathbf{q}),
\end{equation}
where $\hat{T}^2=1$ because we do not explicitly include the spin degree of freedom for the dice lattice. It is straightforward to check that the single-layer dice lattice is time-reversal symmetric with $\hat{T} = \hat{K}$. Since the interlayer coupling Hamiltonians in Eq.~(\ref{Model-h-tun-1-2-3-4}) are real, all stackings considered in this work are time-reversal symmetric.

Finally, let us analyze the inversion symmetry ($\mathcal{P}$ symmetry). This symmetry changes the sign of momentum and interchanges sublattices leaving the Hamiltonian invariant. The operator of the inversion symmetry is $\hat{P}=W\Pi_{\mathbf{q}\to-\mathbf{q}}$, where the matrix $W$ satisfies the following equation:
\begin{align}
\label{Model-symmetry-P}
W \hat{H}(\mathbf{q}) = \hat{H}(-\mathbf{q})W.
\end{align}
In a single-layer dice lattice, the sublattices $A$ and $B$ interchange under the in-plane inversion symmetry. The corresponding matrix $W_0$ is given by the antidiagonal $3\times3$ matrix~\cite{Oriekhov-Gusynin:2018}. For aligned $AA-BB-CC$ and hub-aligned $AB-BA-CC$ stackings, we find the following matrices:
\begin{equation}
\label{Model-symmetry-P-matrix}
W_1=\mathds{1}_2\otimes W_0, \quad
W_2=\tau_x\otimes W_0, \quad
W_0 = \begin{pmatrix}
				0 & 0 & 1\\
				0 & 1 & 0\\
				1 & 0 & 0
			\end{pmatrix}.
\end{equation}
They correspond to the in-plane $W_1$ and full $W_2$ inversion symmetries, respectively. In the former, there is no need to interchange the layers; therefore, in the strict sense, it is not the true inversion symmetry. As with the other discrete symmetries, the aligned $AA-BB-CC$ stacking preserves the inversion symmetry of the dice lattice. As for the hub-aligned $AB-BA-CC$ stacking, the interchange of the layers is equivalent to the rotation by $\pi/3$ with respect to a site $C$. Since the bilayer lattice in the hub-aligned stacking retains the $C_3$ rotation symmetry, it is also invariant with respect to the interchange of the layers. On the other hand, the mixed $AA-BC-CB$ stacking breaks the inversion symmetry. This follows from the fact that the mixed stacking explicitly distinguishes one of the sublattices ($A$ sublattice). It is interesting that the cyclic stacking has no in-plane inversion symmetry; i.e., only the full inversion symmetry with the $W_2$ matrix in Eq.~(\ref{Model-symmetry-P-matrix}) is valid.
The interchange of layers compensates for the change made by the in-plane inversion and restores the cyclic order of atoms.

Let us now identify the inversion centers. We start with the ``in-plane inversion symmetry'': for monolayer and the $AA-BB-CC$ stacking another possible inversion center, in addition to a midpoint between $A$ and $B$ atoms, is a $C$ atom. The $AB-BA-CC$ stacking realizes the same inversion centers. For the mixed $AA-BC-CB$ and cyclic $AB-BC-CA$ stackings, no in-plane inversion symmetry exists. For the ``full inversion symmetry'' the  described centers should be shifted to midpoints between layers. The inversion center for the cyclic $AB-BC-CA$ stacking is, e.g., in the middle of the line connecting the closest $C$ atoms from different layers.

\subsection{Energy spectrum and effective models}
\label{sec:Model-effective}

In this section, we present effective Hamiltonians for bilayer dice models and compare their energy spectra with those of the tight-binding counterparts. In the derivation of the effective models, we follow the standard perturbative approach. The details of the derivation of the effective models can be found in Appendix~\ref{sec:App-Model}. The effective models are derived assuming strong interlayer coupling compared to momenta in the vicinity of the Dirac points, i.e., $g \gg \hbar v_F k$. In addition, in writing linearized effective models, we focus on the $K$ point; the Hamiltonian for the $K^{\prime}$ point can be obtained by replacing $k_x\to-k_x$.

As we show in Figs.~\ref{fig:Model-effective-aligned-TB-energy}--\ref{fig:Model-effective-cyclic-TB-energy}, while the dispersion relation is strongly modified by the inter-layer coupling, the band-crossing points remain gapless~\footnote{
The formation of the gap in dice lattices might be induced, e.g., by interaction effects~\cite{Gorbar-Oriekhov:2021} deviations in the atomic equivalence of three sublattices~\cite{Cunha-Peeters:2021}.}. The inter-layer coupling shifts the points in energy: instead of a doubly degenerate band-crossing point at $g=0$, there are two band-crossing points located at $\pm g$. Effective models are able to capture the most significant features of the dispersion relation in the vicinity of the threefold band-crossing points and provide an analytical description of the deformed flat bands. To simplify the notations, we consider effective models only for the band-crossing point at $g$; the effective models and the energy spectrum for the band-crossing point at $-g$ can be obtained by the replacement $g\to -g$.

It is worth noting that effective models do not always inherit the symmetries of the tight-binding ones. The case in point is the particle-hole symmetry, which requires one to interchange the threefold band-crossing points at $g$ and $-g$ in the tight-binding model. On the other hand, the effective model, which describes only one of these points, may enjoy its own version of the particle-hole symmetry, which reflects the symmetry of the energy spectrum with respect to the band-crossing point. This symmetry is not related to the charge-conjugation symmetry discussed in Sec.~\ref{sec:BL-symmetries} and summarized in Table~\ref{tab:symmetry}.

\subsubsection{Aligned \texorpdfstring{$AA-BB-CC$}{AA-BB-CC} stacking}
\label{sec:Model-effective-aligned}

We start with the simplest aligned $AA-BB-CC$ stacking. The effective Hamiltonian in the vicinity of the $K$ point is
\begin{equation}
\label{Model-effective-aligned-Heff}
H_{\rm eff}^{\rm (a)} = g \mathds{1}_3 + \hbar v_F \left(\mathbf{S}\cdot \mathbf{k}\right).
\end{equation}
As one can see, in the leading nontrivial order in $\hbar v_F k/g$, the effective model for the $AA-BB-CC$ stacking comprises two copies of the single-layer linearized Hamiltonians (the other copy is obtained by replacing $g\to-g$), see Eq.~(\ref{Model-tb-h-lin}), separated by $2g$ in energy. The energy spectrum is given by Eq.~(\ref{Mode-energy-lin}) where the positive and negative branches are shifted by $g$, respectively, i.e.,
\begin{equation}
\label{Model-effective-aligned-energy}
\epsilon_{0} = g, \quad \quad \epsilon_{1} = g + \hbar v_F k,  \quad \mbox{and} \quad \epsilon_{2} = g - \hbar v_F k.
\end{equation}

\begin{figure*}[t]
\centering
\subfigure[]{\includegraphics[width=0.31\textwidth]{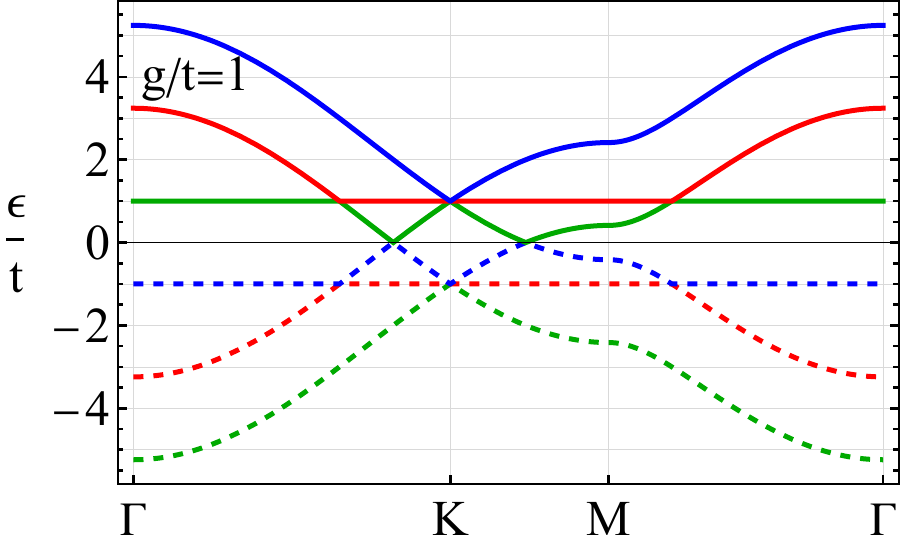}}
\hspace{0.01\textwidth}
\subfigure[]{\includegraphics[width=0.32\textwidth]{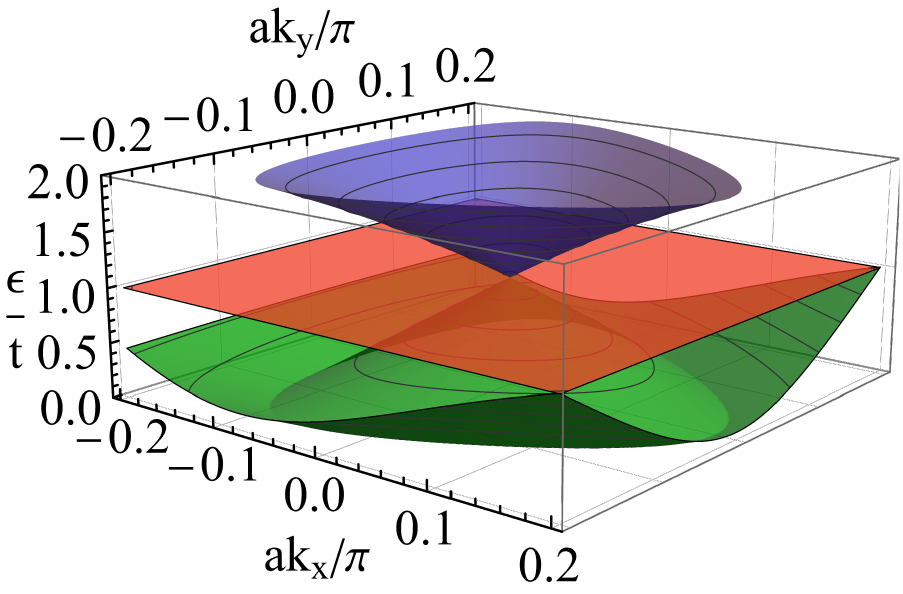}}
\hspace{0.01\textwidth}
\subfigure[]{\includegraphics[width=0.32\textwidth]{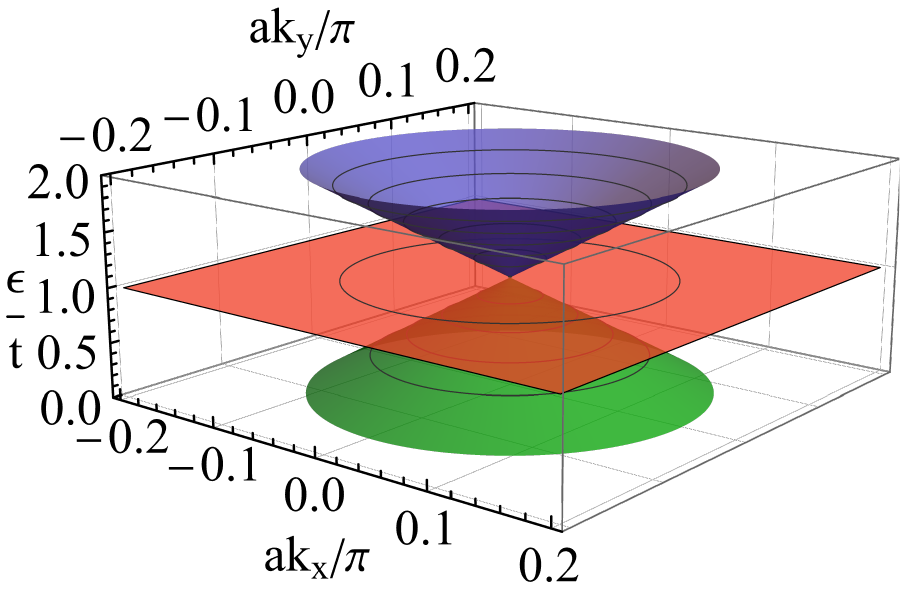}}
\caption{
The energy spectrum of the tight-binding Hamiltonian (\ref{Model-H-total}) for the aligned $AA-BB-CC$ stacking along the $\Gamma-\mbox{K}-\mbox{M}-\Gamma$ line in the Brillouin zone (panel (a)). The tight-binding and effective (see Eq.~(\ref{Model-effective-aligned-energy})), energy spectra at the $K$ point and $\epsilon>0$ are compared in panels (b) and (c), respectively. In all panels, we set $g=t$.
}
\label{fig:Model-effective-aligned-TB-energy}
\end{figure*}

We present the energy dispersion for the tight-binding Hamiltonian (\ref{Model-H-total}) in Fig.~\ref{fig:Model-effective-aligned-TB-energy}(a). The energy spectrum in the vicinity of the $K$ point is compared with that of the effective model in Figs.~\ref{fig:Model-effective-aligned-TB-energy}(b) and \ref{fig:Model-effective-aligned-TB-energy}(c), respectively. Notice that the flat band remains intact. Furthermore, both tight-binding and effective Hamiltonians are particle-hole symmetric as also follows from the symmetry analysis summarized in Table~\ref{tab:symmetry}.

Evidently, the evolution of the energy spectrum with the interlayer coupling constant resembles that in the $A-A$ stacking of bilayer graphene: the band-crossing points in a bilayer dice lattice become separated in energy by $2g$. The energy spectrum at $\epsilon=0$ contains nodal rings around $K$ points. The crosssection of such a nodal ring is shown in Fig.~\ref{fig:Model-effective-aligned-TB-energy}; see also Fig.~\ref{fig:DOS-Spectral-0} for the spectral function.

\subsubsection{Hub-aligned \texorpdfstring{$AB-BA-CC$}{AB-BA-CC} stacking}
\label{sec:Model-effective-hub}

In contrast to the aligned stacking considered in Sec.~\ref{sec:Model-effective-aligned}, the hub-aligned $AB-BA-CC$ stacking requires one to include the second-order in $\hbar v_F k/g$ terms to reproduce an anisotropy of the energy dispersion. The corresponding effective Hamiltonian reads
\begin{widetext}
\begin{equation}
\label{Model-effective-hub-Heff}
H_{\rm eff}^{\rm (h)} = g \mathds{1}_3 + \frac{\hbar v_F}{\sqrt{2}} k_x \left(
                                                            \begin{array}{ccc}
                                                              0 & 1 & 0 \\
                                                              1 & 0 & 1 \\
                                                              0 & 1 & 0 \\
                                                            \end{array}
                                                          \right) +\left(\frac{\hbar v_F}{\sqrt{2}} \right)^2 \frac{k_y^2}{2g}\left(
                                                                     \begin{array}{ccc}
                                                                       1 & 0 & -1 \\
                                                                       0 & 2 & 0 \\
                                                                       -1 & 0 & 1 \\
                                                                     \end{array}
                                                                   \right)
                                                          + \frac{\hbar v_F}{\sqrt{2}} \frac{a}{4} \left(k_y^2 -k_x^2\right) \left(
                                                            \begin{array}{ccc}
                                                              0 & 1 & 0 \\
                                                              1 & 0 & 1 \\
                                                              0 & 1 & 0 \\
                                                            \end{array}
                                                          \right).
\end{equation}
\end{widetext}
The second-order terms are responsible for the asymmetry of the energy spectrum. We have the following energy spectrum in the vicinity of the $K$ point:
\begin{eqnarray}
\label{Model-effective-hub-eps-0}
\epsilon_0 &=& g + \frac{(\hbar v_Fk_y)^2}{2g},\\
\label{Model-effective-hub-eps-1}
\epsilon_{1} &=& g + \frac{(\hbar v_Fk_y)^2}{4g} \nonumber\\
&+& \frac{\hbar v_F}{4g} \sqrt{(\hbar v_F)^2 k_y^4 +g^2 \left[4k_x -a(k_x^2 -k_y^2)\right]^2},\\
\label{Model-effective-hub-eps-2}
\epsilon_{2} &=& g + \frac{(\hbar v_Fk_y)^2}{4g} \nonumber\\
&-& \frac{\hbar v_F}{4g} \sqrt{(\hbar v_F)^2 k_y^4 +g^2 \left[4k_x -a(k_x^2 -k_y^2)\right]^2}.
\end{eqnarray}
If $\hbar v_F/g \gg a$, the terms containing $ak_x^2$ and $ak_y^2$, i.e., the last term in Eq.~(\ref{Model-effective-hub-Heff}) can be neglected. Then, the energy spectrum in Eqs.~(\ref{Model-effective-hub-eps-1})--(\ref{Model-effective-hub-eps-2}) corresponds to a particle-hole asymmetric version of the semi-Dirac spectrum~\cite{Hasegawa-Kohmoto:2006} in which the dispersion relation is linear in one direction and quadratic in the other. The particle-hole symmetry breakdown around each of the band-crossing points is quantified by momentum-dependent $\sim (\hbar v_Fk_y)^2/g$ term.

\begin{figure*}[t]
\centering
\subfigure[]{\includegraphics[width=0.31\textwidth]{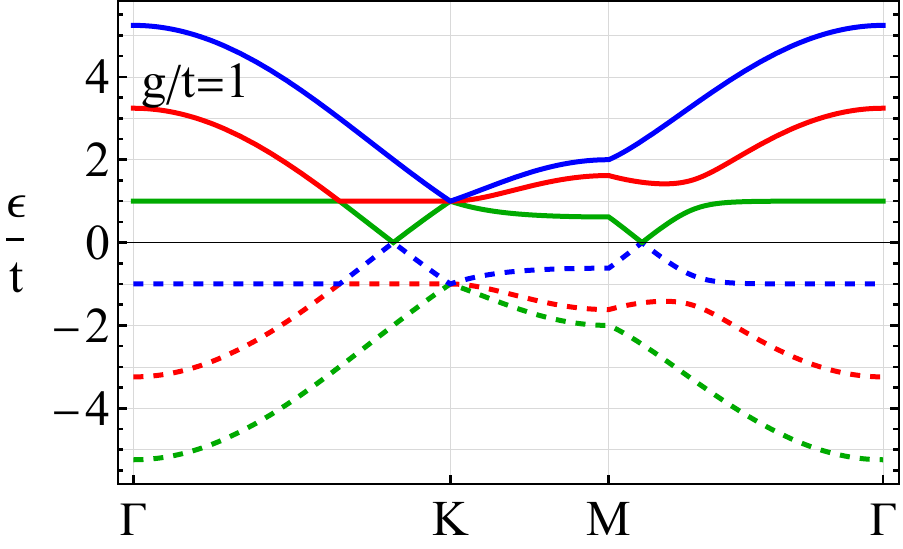}}
\hspace{0.01\textwidth}
\subfigure[]{\includegraphics[width=0.32\textwidth]{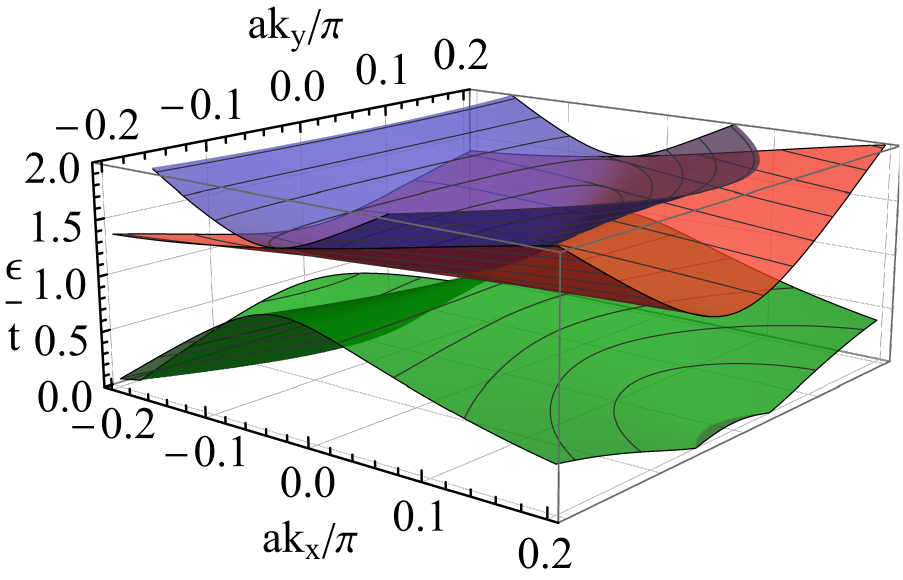}}
\hspace{0.01\textwidth}
\subfigure[]{\includegraphics[width=0.32\textwidth]{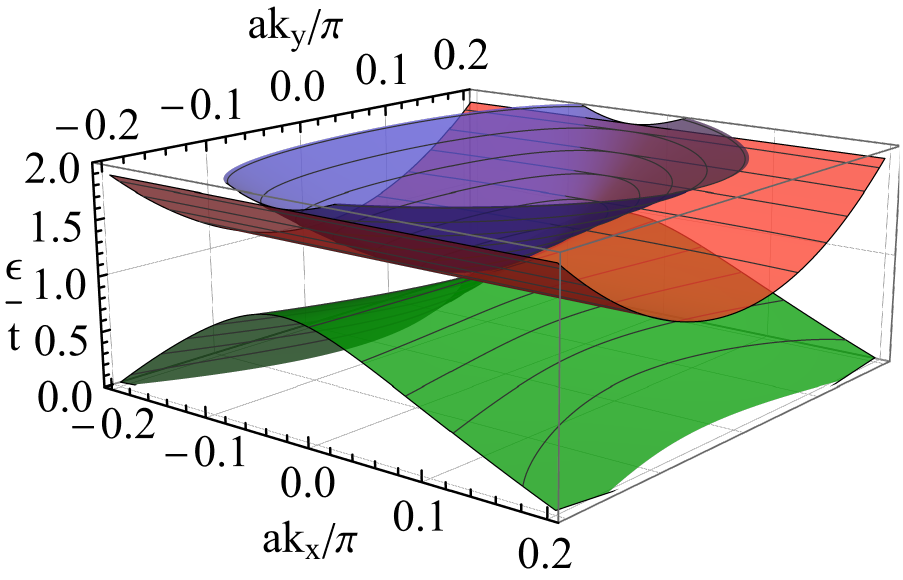}}
\caption{
The energy spectrum of the tight-binding Hamiltonian (\ref{Model-H-total}) for the hub-aligned $AB-BA-CC$ stacking along the $\Gamma-\mbox{K}-\mbox{M}-\Gamma$ line in the Brillouin zone (panel (a)). The tight-binding and effective, see Eqs.~(\ref{Model-effective-hub-eps-0})--(\ref{Model-effective-hub-eps-2}), energy spectra at the $K$ point and $\epsilon>0$ are compared in panels (b) and (c), respectively. In all panels, we set $g=t$.
}
\label{fig:Model-effective-hub-TB-energy}
\end{figure*}

We present the energy dispersion for the tight-binding Hamiltonian (\ref{Model-H-total}) in Fig.~\ref{fig:Model-effective-hub-TB-energy}(a). The energy spectrum in the vicinity of the $K$ point is compared with that of the effective model in Figs.~\ref{fig:Model-effective-hub-TB-energy}(b) and \ref{fig:Model-effective-hub-TB-energy}(c), respectively. The spectrum is clearly anisotropic with a linear dispersion relation along $k_x$ and the quadratic one along $k_y$. Furthermore, the particle-hole symmetry is absent for the effective model (i.e., the bands in the vicinity of the band-crossing points are asymmetric at $\epsilon = \pm g$) but is present in the tight-binding one; see Fig.~\ref{fig:Model-effective-hub-TB-energy}(a) and Table~\ref{tab:symmetry}. It is interesting to notice also that the energy spectrum for the hub-aligned $AB-BA-CC$ stacking retains some features of the spectrum of the aligned $AA-BB-CC$ stacking; namely, the band remains flat along certain directions ($k_y$) [cf. Figs.~\ref{fig:Model-effective-aligned-TB-energy}(a) and \ref{fig:Model-effective-hub-TB-energy}(a)]. In addition, the bands at $\epsilon=0$ intersect along lines in momentum space rather than form nodes; see also Fig.~\ref{fig:DOS-Spectral-0} for the spectral function.

\subsubsection{Mixed \texorpdfstring{$AA-BC-CB$}{AA-BC-CB} stacking}
\label{sec:Model-effective-mixed}

In the case of the mixed $AA-BC-CB$ stacking with the coupling Hamiltonian defined by $H_{\rm c}^{\rm (m)}$ in Eq.~(\ref{Model-h-tun-1-2-3-4}), we derive the following effective Hamiltonian:
\begin{widetext}
\begin{eqnarray}
\label{Model-effective-mixed-Heff}
H_{\rm eff}^{\rm (m)} &=& g \mathds{1}_3 + \frac{\hbar v_F}{2\sqrt{2}} \left(
                                                            \begin{array}{ccc}
                                                              0 & 2k_x & k_{-} \\
                                                              2k_x & 0 & k_{-} \\
                                                              k_{+} & k_{+} & 0 \\
                                                            \end{array}
                                                          \right) + \left(\frac{\hbar v_F}{4} \right)^2 \frac{1}{g}\left(
                                        \begin{array}{ccc}
                                          k_x^2 +5k_y^2 & 0 & 0 \\
                                          0 & k_x^2 +5k_y^2 & 0 \\
                                          0 & 0 & 2k^2 \\
                                        \end{array}
                                      \right) \nonumber\\
&-& \left(\frac{\hbar v_F}{4} \right)^2 \frac{1}{g}\left(
                                        \begin{array}{ccc}
                                          0 & k^2 & 2ik_y k_{-} \\
                                          k^2 & 0 & 2ik_y k_{-} \\
                                          -2ik_y k_{+} & -2ik_y k_{+} & 0 \\
                                        \end{array}
                                      \right)
- \frac{\hbar v_F a}{8\sqrt{2}} \left(
                                        \begin{array}{ccc}
                                          0 & 2(k_x^2-k_y^2) & k_{+}^2 \\
                                          2(k_x^2-k_y^2) & 0 & k_{+}^2 \\
                                          k_{-}^2 & k_{-}^2 & 0 \\
                                        \end{array}
                                      \right).
\end{eqnarray}
\end{widetext}

The energy spectrum up to the second order in momentum is quite cumbersome. Therefore, we leave the second-order terms only in the $\epsilon_0$ branch where they are crucial to describe the anisotropy and provide leading order corrections at $k_x=0$. For other branches, the second-order terms can be neglected compared to the leading-order linear terms. Therefore, we have
\begin{eqnarray}
\label{Model-effective-mixed-eps-0}
\epsilon_0 &=& g - \frac{\hbar v_F}{\sqrt{2}} k_x + \frac{(\hbar v_F)^2}{8g} \left(k_x^2 +3k_y^2\right) \nonumber\\
&+& \frac{\hbar v_F}{4\sqrt{2}} a \left(k_x^2 -k_y^2\right),\\
\label{Model-effective-mixed-eps-1}
\epsilon_1 &=& g + \frac{\hbar v_F}{2\sqrt{2}} k_x + \frac{\hbar v_F}{2\sqrt{2}} \sqrt{3k_x^2 +2k_y^2},\\
\label{Model-effective-mixed-eps-2}
\epsilon_2 &=& g + \frac{\hbar v_F}{2\sqrt{2}} k_x - \frac{\hbar v_F}{2\sqrt{2}} \sqrt{3k_x^2 +2k_y^2}.
\end{eqnarray}

\begin{figure*}[t]
\centering
\subfigure[]{\includegraphics[width=0.31\textwidth]{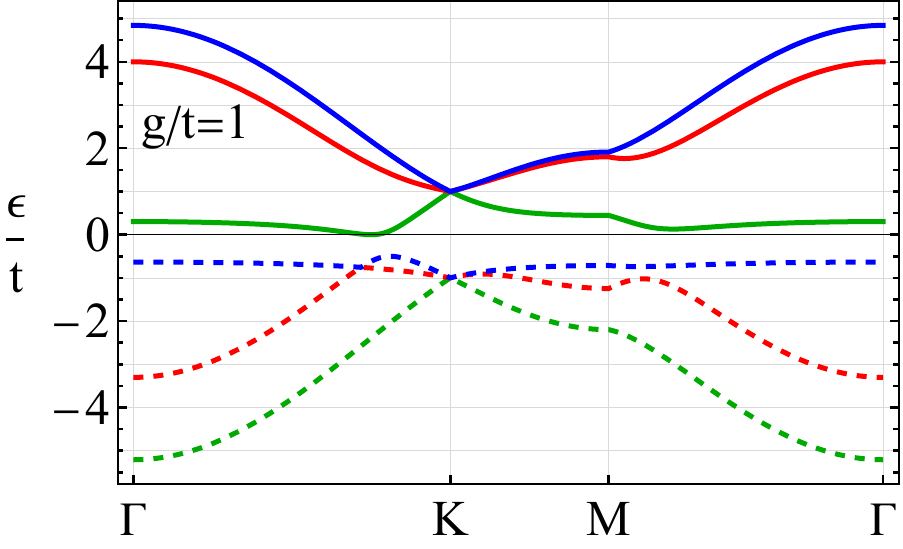}}
\hspace{0.01\textwidth}
\subfigure[]{\includegraphics[width=0.32\textwidth]{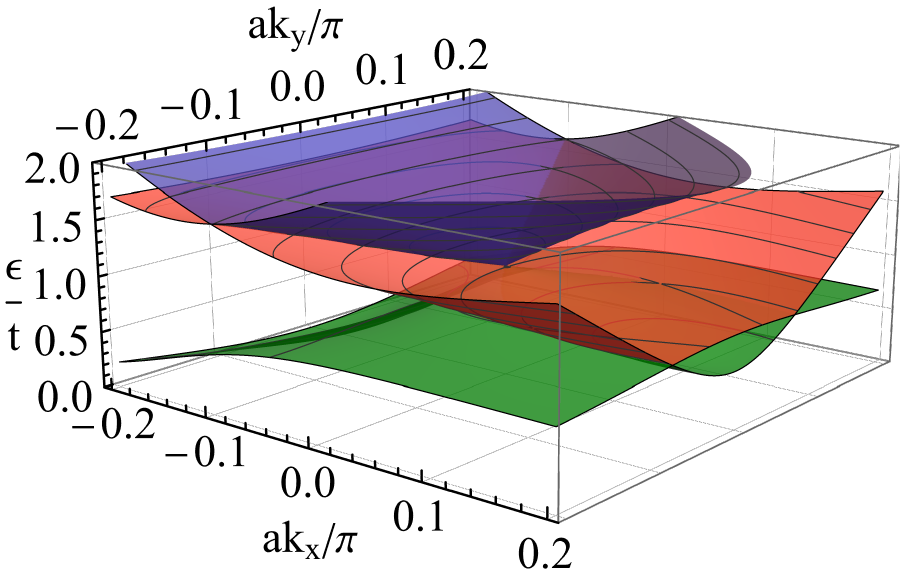}}
\hspace{0.01\textwidth}
\subfigure[]{\includegraphics[width=0.32\textwidth]{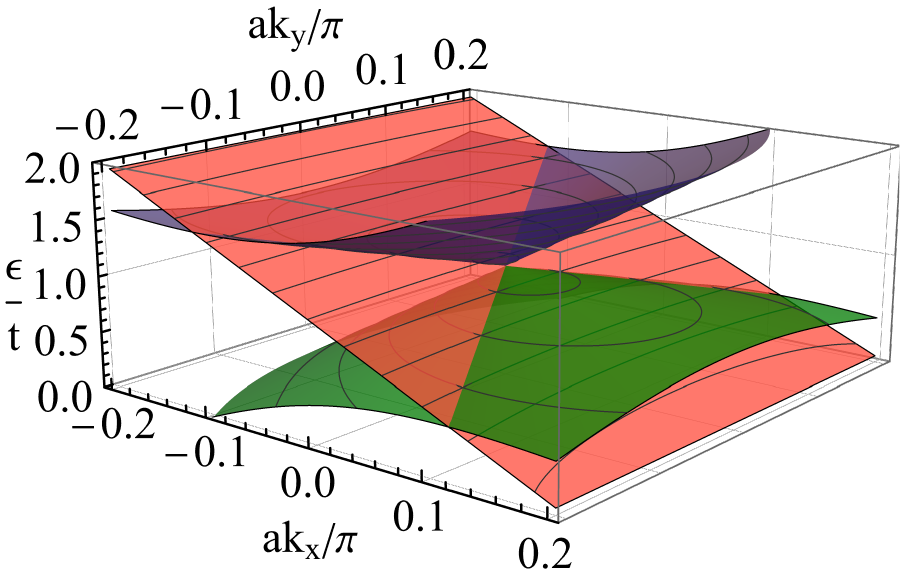}}
\caption{
The energy spectrum of the tight-binding Hamiltonian (\ref{Model-H-total}) for the mixed $AA-BC-CB$ stacking along the $\Gamma-\mbox{K}-\mbox{M}-\Gamma$ line in the Brillouin zone (panel (a)). The tight-binding and effective, see Eqs.~(\ref{Model-effective-mixed-eps-0})--(\ref{Model-effective-mixed-eps-2}), energy spectra at the $K$ point and $\epsilon>0$ are compared in panels (b) and (c), respectively.
In all panels, we set $g=t$.
}
\label{fig:Model-effective-mixed-TB-energy}
\end{figure*}

We present the energy dispersion for the tight-binding Hamiltonian (\ref{Model-H-total}) with the coupling Hamiltonian $H_{\rm c}^{\rm (m)}$ defined in Eq.~(\ref{Model-h-tun-1-2-3-4}) in Fig.~\ref{fig:Model-effective-mixed-TB-energy}(a). The tight-binding energy spectrum in the vicinity of the $K$ point is compared with that of the effective model (\ref{Model-effective-mixed-Heff}) in Figs.~\ref{fig:Model-effective-mixed-TB-energy}(b) and \ref{fig:Model-effective-mixed-TB-energy}(c), respectively. As one can see, dispersive Dirac-like bands become anisotropic. Furthermore, as in the case of the hub-aligned $AB-BA-CC$ stacking, the additional band is no longer flat but acquires a noticeable anisotropic dispersion along all directions. Another noticeable feature of the spectrum is the absence of particle-hole symmetry in the tight-binding and effective models. That is, the energy spectrum is asymmetric both at $\epsilon=0$ and $\epsilon=\pm g$. This is qualitatively different from the hub-aligned $AB-BA-CC$ stacking where the energy spectrum is symmetric with respect to $\epsilon=0$; cf. Figs.~\ref{fig:Model-effective-hub-TB-energy}(a) and \ref{fig:Model-effective-mixed-TB-energy}(a). The lack of particle-hole symmetry is directly related to the lack of the charge-conjugation symmetry of the tight-binding Hamiltonian; see Table~\ref{tab:symmetry}.

Compared to the aligned and hub-aligned stackings, the energy spectrum at $\epsilon=0$ is drastically different. As is evident from Fig.~\ref{fig:Model-effective-mixed-TB-energy}(a), the bands no longer cross. Still, one of the bands may attain zero values; see the solid green line in Fig.~\ref{fig:Model-effective-mixed-TB-energy}(a).

\subsubsection{Cyclic \texorpdfstring{$AB-BC-CA$}{AB-BC-CA} stacking}
\label{sec:Model-effective-cyclic}

The effective linearized Hamiltonian for the cyclic $AB-BC-CA$ stacking reads
\begin{equation}
\label{Model-effective-cyclic-Heff}
H_{\rm eff}^{\rm (c)} = g \mathds{1}_3 + \frac{\hbar v_F}{2\sqrt{2}} \left(
                                                            \begin{array}{ccc}
                                                              0 & k_{-} & k_{+} \\
                                                              k_{+} & 0 & 2k_{-} \\
                                                              k_{-} & 2k_{+} & 0 \\
                                                            \end{array}
                                                          \right).
\end{equation}
The energy spectrum is determined by the following third-order equation:
\begin{equation}
\label{Model-effective-cyclic-eps-third}
\left(\epsilon - g\right)^3 - A_1\left(\epsilon - g\right) +A_2 =0,
\end{equation}
where
\begin{equation}
\label{Model-effective-cyclic-eps-third-sol-A1-A2}
A_1 = \frac{27}{8} (atk)^2 \quad \mbox{and} \quad  A_2 = \frac{27}{16} (at)^3 k_x \left(k_x^2-3k_y^2\right).
\end{equation}
The solutions to Eq.~(\ref{Model-effective-cyclic-eps-third}) are
\begin{eqnarray}
\label{Model-effective-cyclic-eps-0}
\epsilon_0 &=& g +2 \sqrt{\frac{A_1}{3}} \cos{\left[\frac{1}{3} \mbox{arccos}{\left(\frac{3A_2}{2A_1} \sqrt{\frac{3}{A_1}}\right)} -\frac{2\pi}{3}\right]} \nonumber\\
&=& g + \hbar v_F k \cos{\left\{\frac{1}{3} \mbox{arccos}{\left[\frac{\cos{(3\varphi)}}{\sqrt{2}}\right]} -\frac{2\pi}{3}\right\}},\\
\label{Model-effective-cyclic-eps-1}
\epsilon_1 &=& g +2 \sqrt{\frac{A_1}{3}} \cos{\left[\frac{1}{3} \mbox{arccos}{\left(\frac{3A_2}{2A_1} \sqrt{\frac{3}{A_1}}\right)}\right]} \nonumber\\
&=& g + \hbar v_F k \cos{\left\{\frac{1}{3} \mbox{arccos}{\left[\frac{\cos{(3\varphi)}}{\sqrt{2}}\right]} \right\}},\\
\label{Model-effective-cyclic-eps-2}
\epsilon_2 &=& g +2 \sqrt{\frac{A_1}{3}} \cos{\left[\frac{1}{3} \mbox{arccos}{\left(\frac{3A_2}{2A_1} \sqrt{\frac{3}{A_1}}\right)} -\frac{4\pi}{3}\right]} \nonumber\\
&=& g + \hbar v_F k \cos{\left\{\frac{1}{3} \mbox{arccos}{\left[\frac{\cos{(3\varphi)}}{\sqrt{2}}\right]} -\frac{4\pi}{3}\right\}}.
\end{eqnarray}
In the second expressions in Eqs.~(\ref{Model-effective-cyclic-eps-0})--(\ref{Model-effective-cyclic-eps-2}), we used the polar coordinate system with $\left\{k_x, k_y\right\} = k \left\{\cos{\varphi}, \sin{\varphi}\right\}$.

\begin{figure*}[t]
\centering
\subfigure[]{\includegraphics[width=0.31\textwidth]{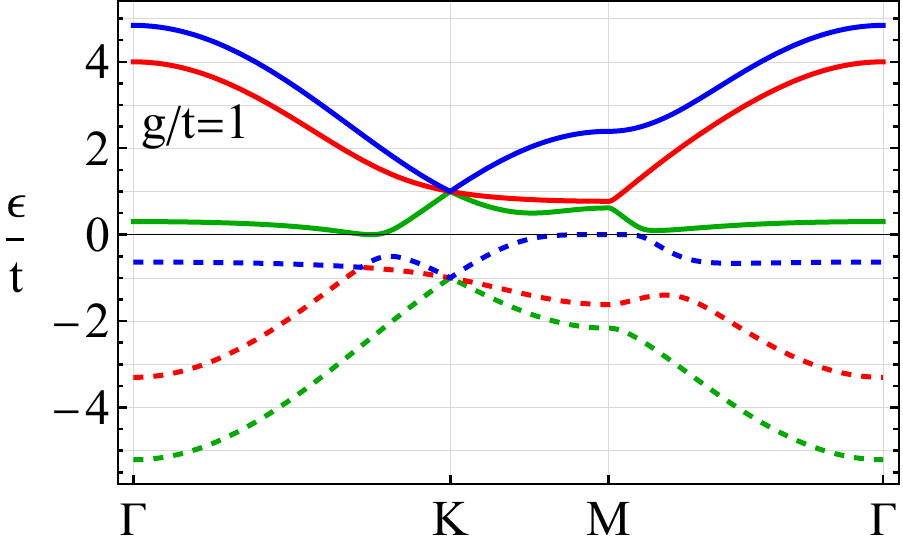}}
\hspace{0.01\textwidth}
\subfigure[]{\includegraphics[width=0.32\textwidth]{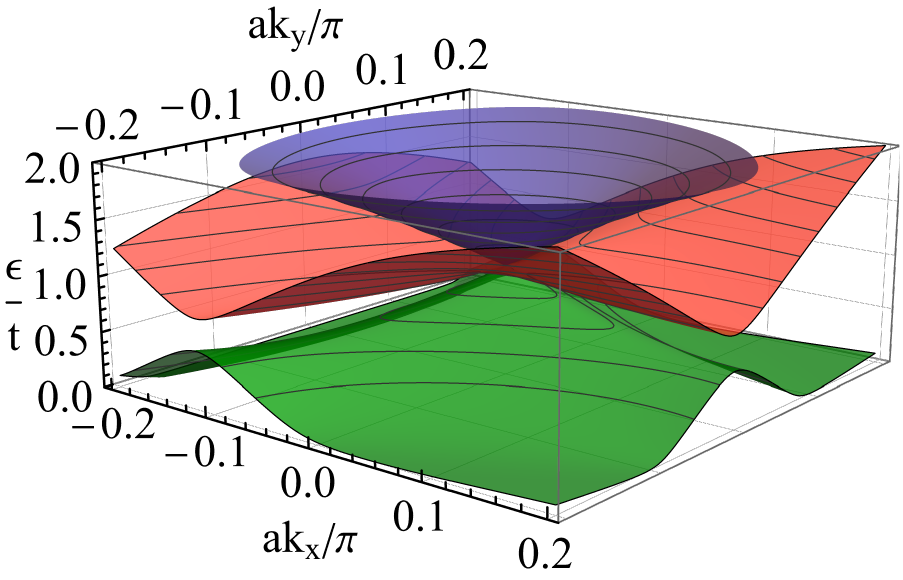}}
\hspace{0.01\textwidth}
\subfigure[]{\includegraphics[width=0.32\textwidth]{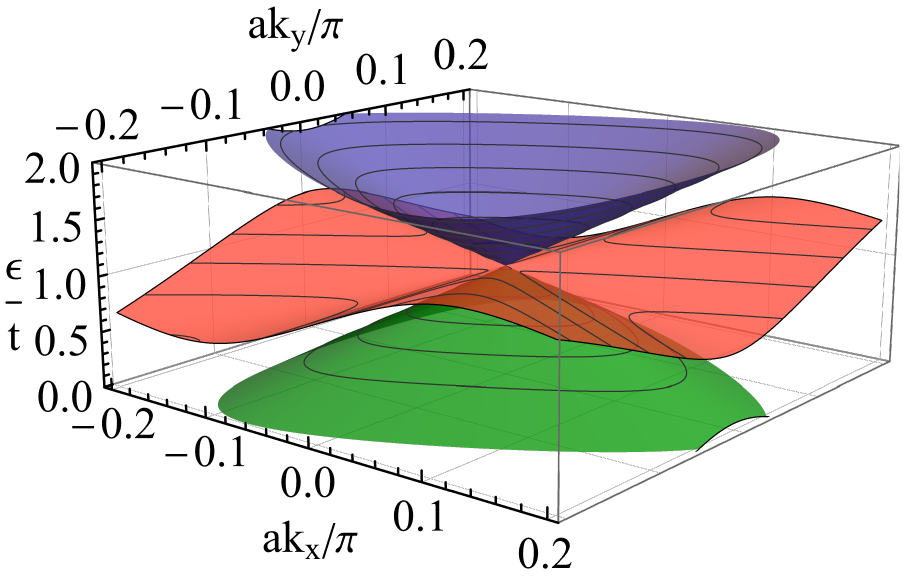}}
\caption{
The energy spectrum of the tight-binding Hamiltonian (\ref{Model-H-total}) for the cyclic $AB-BC-CA$ stacking along the $\Gamma-\mbox{K}-\mbox{M}-\Gamma$ line in the Brillouin zone (panel (a)). The tight-binding and effective, see Eqs.~(\ref{Model-effective-cyclic-eps-0})--(\ref{Model-effective-cyclic-eps-2}), energy spectra at the $K$ point and $\epsilon>0$ are compared in panels (b) and (c), respectively.
In all panels, we set $g=t$.
}
\label{fig:Model-effective-cyclic-TB-energy}
\end{figure*}

We present the energy dispersion for the tight-binding Hamiltonian (\ref{Model-H-total}) with the coupling Hamiltonian $H_{\rm c}^{\rm (c)}$, see Eq.~(\ref{Model-h-tun-1-2-3-4}), in Fig.~\ref{fig:Model-effective-cyclic-TB-energy}(a). The tight-binding energy spectrum in the vicinity of the $K$ point is compared with that of the effective model in Figs.~\ref{fig:Model-effective-cyclic-TB-energy}(b) and \ref{fig:Model-effective-cyclic-TB-energy}(c), respectively. As one can see, both dispersive and flat bands become corrugated due to the interlayer coupling. The corrugation has $C_3$ symmetry; see also Eqs.~(\ref{Model-effective-cyclic-eps-0})--(\ref{Model-effective-cyclic-eps-2}). Despite being linear in momentum, the effective model captures the main features of the energy spectrum reasonably well. The particle-hole symmetry is absent both in tight-binding and effective models; i.e., the energy spectrum is asymmetric both with respect to $\epsilon =0$ and $\epsilon=\pm g$ (see also Table~\ref{tab:symmetry}).

The low-energy spectrum $|\epsilon|/t\ll1$ is similar to that for the mixed stacking, but shows a semimetallic rather than semiconductor-like behavior. The electron and hole bands are located in different parts of the Brillouin zone, see Fig.~\ref{fig:Model-effective-cyclic-TB-energy}(a). The electron and hole pockets form a rather intricate kagome pattern at $\epsilon=0$, see also Fig.~\ref{fig:DOS-Spectral-0}(d).

\section{Density of states and spectral function}
\label{sec:DOS}

In this section, we discuss the spectral function and the DOS for the bilayer dice lattices. To start with, we introduce the Green function in the momentum space,
\begin{equation}
\label{DOS-G-def}
G(\omega \pm i 0; \mathbf{k})  = \frac{i}{\hbar \omega -\mu -H(\mathbf{k}) \pm i0},
\end{equation}
where $H(\mathbf{k})$ is the Hamiltonian (effective or tight-binding), $\mu$ is the Fermi energy, and signs $\pm$ define the retarded ($+$) and advanced ($-$) Green functions. By using the Green function (\ref{DOS-G-def}), we define the spectral function
\begin{equation}
\label{DOS-A-def}
A(\omega; \mathbf{k}) =\frac{1}{2\pi} \left[ G(\omega+i 0; \mathbf{k}) -G(\omega-i 0; \mathbf{k}) \right] \Big|_{\mu=0}.
\end{equation}
While the complete information about the spectral properties is provided by the spectral function $A(\omega; \mathbf{k})$, another useful quantity measured in, e.g., scanning tunneling spectroscopy experiments, is the DOS $\nu(\omega)$ defined as
\begin{equation}
\label{DOS-DOS-def}
\nu(\omega)  = \int \frac{d^2k}{(2\pi)^2} \mbox{tr}{\left\{A(\omega; \mathbf{k})\right\}},
\end{equation}
where the integration proceeds over the Brillouin zone if the tight-binding Hamiltonian is used.

The explicit form of the Green and spectral functions is rather cumbersome even for the effective Hamiltonians. Only the case of the aligned $AA-BB-CC$ stacking is relatively simple because it corresponds to two copies of a single-layer dice model; see, e.g., Ref.~\cite{Gorbar-Oriekhov:2021} for the expressions for the Green function. The corresponding DOS for the effective model reads as
\begin{equation}
\label{DOS-DOS-ii}
\nu^{\rm (a)}(\omega)  = \frac{1}{2\pi (\hbar v_F)^2} \left[\frac{\Lambda^2}{2}\delta{\left(\hbar \omega - g\right)}
+\left|\hbar \omega - g\right| \right],
\end{equation}
where $\Lambda$ is the energy cutoff. The first term in Eq.~(\ref{DOS-DOS-ii}) is related to the flat band contribution and the second term has the same form as the DOS in monolayer graphene. The DOS (\ref{DOS-DOS-ii}) is essentially the same as for the single-layer dice model~\cite{Bercioux-Haeusler:2009}.

\begin{figure*}[!ht]
\centering
\subfigure[\empty]{\includegraphics[width=0.95\textwidth]{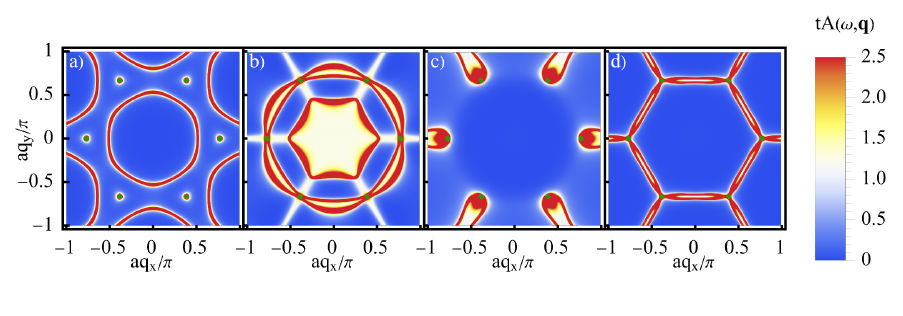}}
\hspace{0.01\textwidth}
\subfigure[\empty]{\includegraphics[width=0.95\textwidth]{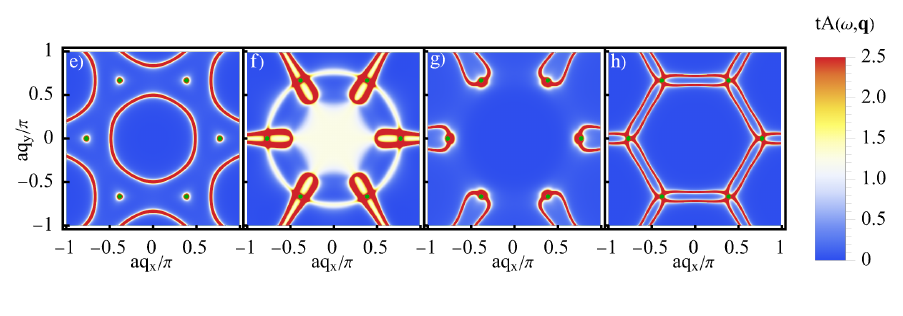}}
\caption{
The spectral functions in the vicinity of the band-crossing points. The upper and lower panels correspond to $\hbar \omega/t=0.9$ and $\hbar \omega/t=1.1$, respectively. The columns represent the results for the aligned $AA-BB-CC$ [panels (a) and (e)], hub-aligned $AB-BA-CC$ [panels (b) and (f)], mixed $AA-BC-CB$ [panels (c) and (g)], and cyclic $AB-BC-CA$ [panels (d) and (h)] stackings. In all panels, we set $g=t$. We use tight-binding models with the spectral function defined in Eq.~(\ref{DOS-A-def}) and introduce the phenomenological broadening $\Gamma = 0.05\,t$ by replacing $i0\to i\Gamma$ in the Green function.
}
\label{fig:DOS-Spectral}
\end{figure*}

The spectral functions for the four stackings are presented in Fig.~\ref{fig:DOS-Spectral}. We focus on the energies in the vicinity of the band-touching points and set $g/t=1$. As one can see, there is a rather intricate pattern where the energy spectrum is evidently asymmetric with respect to the band-crossing points for all stackings except the aligned one; see Figs.~\ref{fig:DOS-Spectral}(a) and \ref{fig:DOS-Spectral}(e). The shape of the spectrum is noticeably different for the energies below and above the band crossing point for the hub-aligned $AB-BA-CC$ stacking which is related to its peculiar particle-hole asymmetric semi-Dirac spectrum; see Fig.~\ref{fig:Model-effective-hub-TB-energy} as well as Figs.~\ref{fig:DOS-Spectral}(b) and \ref{fig:DOS-Spectral}(f). The Dirac point intersected with the tilted band can be inferred from Figs.~\ref{fig:DOS-Spectral}(c) and \ref{fig:DOS-Spectral}(g) for the mixed $AA-BC-CB$ stacking. Finally, the asymmetry is related primarily to the additional $C_3$-corrugated band for the cyclic $AB-BC-CA$ stacking; see Figs.~\ref{fig:DOS-Spectral}(d) and \ref{fig:DOS-Spectral}(h).

\begin{figure*}[t]
\includegraphics[width=0.99\textwidth]{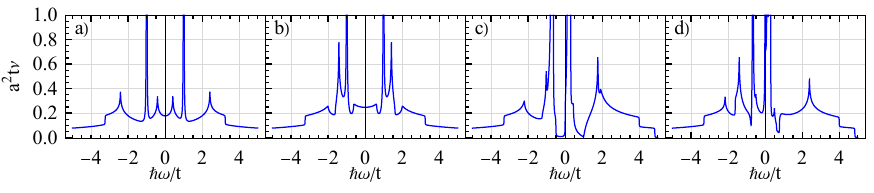}
\caption{
The density of states for bilayer dice lattices in four stackings: aligned $AA-BB-CC$ (panel (a)), hub-aligned $AB-BA-CC$ (panel (b)), mixed $AA-BC-CB$ (panel (c)), and cyclic $AB-BC-CA$ (panel (d)). For all stackings, we employed the tight-binding model with the spectral function defined in Eq.~(\ref{DOS-A-def}) and introduced the phenomenological broadening $\Gamma/t = 0.005$ by replacing $i0\to i\Gamma$ in the Green function.
}
\label{fig:DOS-DOS-all}
\end{figure*}

By integrating the spectral function over the Brillouin zone, we obtain the DOS in Fig.~\ref{fig:DOS-DOS-all}. As expected, the DOS has the simplest structure for the aligned $AA-BB-CC$ stacking and reveals the peaks corresponding to the flat bands at $\hbar \omega = \pm g$, see Eq.~(\ref{DOS-DOS-ii}), as well as two sets of smaller peaks corresponding to the van Hove singularities; see Fig.~\ref{fig:DOS-DOS-all}(a). A similar structure of the DOS with well-pronounced peaks at $\hbar\omega=\pm g$ is observed for the hub-aligned $AB-BA-CC$ stacking with, however, different locations of the van Hove singularities; see Fig.~\ref{fig:DOS-DOS-all}(b). The DOS for the mixed $AA-BC-CB$ and cyclic $AB-BC-CA$ stackings has a rather complicated structure with several peaks and absent particle-hole symmetry; see Figs.~\ref{fig:DOS-DOS-all}(c) and \ref{fig:DOS-DOS-all}(d). In both cases, there are peaks near $\hbar\omega=0$ and $\hbar \omega =-g$, while the DOS at $\hbar \omega =g$ is suppressed. Unlike the aligned and hub-aligned stackings where the peaks at $\hbar \omega = \pm g$ are related to flat or partially flat (having a softer dispersion relation along one of the directions) bands, all peaks for the mixed and cyclic stackings correspond to the extrema in the energy spectrum. Another difference between these stackings is related to the particle-hole symmetry. The DOSs for the aligned and hub-aligned stackings are symmetric with respect to both $\epsilon=0$ and $\epsilon=\pm g$; see Appendix~\ref{sec:App-spectral-low} for the results at larger $g$ where the approximate symmetry becomes evident. On the other hand, the peaks in the DOS for the mixed $AA-BC-CB$ and cyclic $AB-BC-CA$ stackings are always asymmetric; this result persists also for larger $g$ (see Appendix~\ref{sec:App-spectral-low}).

\section{Summary}
\label{sec:Summary}

In this work, we introduced and classified the nonequivalent vertically aligned commensurate stackings for bilayer dice ($\mathcal{T}_3$) lattices. These four stackings are the aligned $AA-BB-CC$, hub-aligned $AA-BC-CB$, mixed $AB-BA-CC$, and cyclic $AB-BC-CA$ stacking. Other stackings are either equivalent, nonvertically aligned, or noncommensurate. We found that the bilayer dice model demonstrates a unique energy spectrum for each of the stackings.

In all stackings considered in this work, three energy bands intersect at the $K$ and $K^{\prime}$ points; the band-crossing points are separated in energy with the separation determined by the interlayer coupling constant $g$. The spectrum of the aligned $AA-BB-CC$ stacking resembles that of two copies of the single-layer dice model and contains Dirac points intersected by a completely flat in the whole Brillouin zone band; see Fig.~\ref{fig:Model-effective-aligned-TB-energy}. The hub-aligned $AB-BA-CC$ stacking allows one to realize the semi-Dirac spectrum in the vicinity of the band-crossing points, for which the dispersion relation is quadratic in one direction and linear in the other; see Fig.~\ref{fig:Model-effective-hub-TB-energy}. An unusual spectrum composed of a Dirac point intersected by a tilted anisotropic band occurs for the mixed $AA-BC-CB$ stacking; see Fig.~\ref{fig:Model-effective-mixed-TB-energy}. Somewhat similar to the case of the hub-aligned $AB-BA-CC$ stacking, all bands have a semi-Dirac spectrum. Finally, the cyclic $AB-BC-CA$ stacking realizes an anisotropic energy spectrum with a $C_3$-corrugated additional band intersecting the Dirac point; see Fig.~\ref{fig:Model-effective-cyclic-TB-energy}. The low-energy spectrum, i.e., at $|\epsilon|\ll g$, also depends on the stackings and shows nodal-line crossings (aligned and hub-aligned stackings), semiconductor-like behavior (mixed stacking), or semimetallic-like shape (cyclic stacking) in which conduction and valence bands acquire the same energy but are separated in the Brillouin zone; see Fig.~\ref{fig:Model-effective-cyclic-TB-energy}.
Therefore, similar to multilayer graphene structures, a multilayer dice lattice also holds a potential to be a flexible platform for realizing different types of quasiparticle spectra.

To clarify the shape of the energy spectrum in the vicinity of the threefold band-crossing points and set the stage for analytical calculations, we derived effective models. The corresponding Hamiltonians are given in Eqs.~(\ref{Model-effective-aligned-Heff}), (\ref{Model-effective-hub-Heff}), (\ref{Model-effective-mixed-Heff}), and (\ref{Model-effective-cyclic-Heff}). The energy spectrum of these models captures the main features of the tight-binding spectrum such as the anisotropy of the dispersion relation. Furthermore, the effective models allow us to introduce effective particle-hole symmetry with respect to the band-crossing points. In particular, the aligned $AA-BB-CC$ stacking shows particle-hole symmetry for both tight-binding and effective (i.e., with respect to each of the band-crossing points) models. While the tight-binding model is particle-hole symmetric, there is no particle-hole symmetry for the effective model of the hub-aligned $AB-BA-CC$ stacking. For the other two stackings, i.e., the mixed $AA-BC-CB$ and cyclic $AB-BC-CA$ ones, both tight-binding and effective models are particle-hole asymmetric. The derived effective models might be useful in various applications including the studies of transport, collective modes, edge states, etc.

We used the obtained tight-binding models to calculate the spectral function and the DOS in Sec.~\ref{sec:DOS}; see Figs.~\ref{fig:DOS-Spectral} and \ref{fig:DOS-DOS-all}. The spectral function provides access to the cross-sections of the energy dispersion, which could become rather intricate for certain stackings. The intricate band structure of the bilayer dice model also has a direct manifestation in the DOS. In particular, the flat band of the aligned $AA-BB-CC$ stacking leads to peaks corresponding to the band-crossing points. The peaks are also observed for the hub-aligned $AB-BA-CC$ stacking due to a soft, but not exactly flat, dispersion relation of the additional band. On the other hand, the DOS of the mixed $AA-BC-CB$ and cyclic $AB-BC-CA$ stackings is dominated by the van Hove singularities related to the features of the spectrum away from the band-crossing points. In solid-state realizations of the dice lattice, the spectral function and the DOS can be probed via angle-resolved photoemission and scanning tunneling spectroscopy experiments.

In the derivation of bilayer dice models, we have made a few simplifying assumptions related to the structure of the lattice and the coupling Hamiltonian. First, we considered only commensurate stackings where sublattices of both layers are vertically aligned. In writing the coupling Hamiltonians (\ref{Model-h-tun-1-2-3-4}), only the nearest-neighbor hopping and equal coupling constants for all sites were assumed. The breakdown of the symmetry between the $A$ and $B$ sublattices might lead to a few additional stackings. It would be also interesting to investigate which of the proposed stackings is the most energetically favorable. These studies are beyond this work and will be reported elsewhere. Finally, we notice that the rich energy spectrum and nontrivial DOS promise unusual optical responses of bilayer dice lattices. The studies of the optical response are presented in our work~\cite{SOG:part2-2023}.

\begin{acknowledgments}
P.O.S. acknowledges support through the Yale Prize Postdoctoral Fellowship in Condensed Matter Theory. D.O.O. acknowledges the support from the Netherlands Organization for Scientific Research (NWO/OCW) and from the European Research Council (ERC) under the European Union's Horizon 2020 research and innovation program.
\end{acknowledgments}


\appendix

\section{Derivation of the effective model}
\label{sec:App-Model}

In this appendix, we discuss the derivation of the effective Hamiltonians presented in Sec.~\ref{sec:Model-effective}; see also Ref.~\cite{McCann-Koshino:2012} for a similar discussion for bilayer graphene. We focus on the dynamics in the vicinity of band crossing points, i.e., at $|\epsilon| \approx g$. Then, the off-diagonal terms in the Hamiltonian (\ref{Model-H-total}) with the coupling Hamiltonians defined in Eq.~(\ref{Model-h-tun-1-2-3-4}) are assumed to be large compared to the diagonal ones, i.e., $g/(\hbar v_F q) \gg1$. In this case, it is convenient to transform the full Hamiltonian (\ref{Model-H-total}) into a new basis where the part of the Hamiltonian responsible for the interlayer coupling, i.e., the Hamiltonian (\ref{Model-H-total}) with $H(\mathbf{q})=0$, is diagonal. This allows us to separate the low- and high-energy (with respect to the band-crossing point at $\epsilon = g$) parts of the full Hamiltonian as
\begin{equation}
\label{App-Model-Heff-def}
H = \left(
      \begin{array}{cc}
        h_{g} & u \\
        u^{\dag} & h_{-g} \\
      \end{array}
    \right),
\end{equation}
where $h_{g}$ and $h_{-g}$ describe the states in the vicinity of the crossing points at $\epsilon= g$ and $\epsilon= -g$, respectively. The coupling between them is denoted by $u$. Now, the off-diagonal terms in Eq.~(\ref{App-Model-Heff-def}) are small compared to the diagonal ones and can apply the standard perturbative approach. It is convenient to separate
\begin{equation}
\label{App-Model-Heff-hL-separate}
h_{g} = h_{g}^{(0)}+\delta h_{g} \quad \mbox{and} \quad h_{-g} = h_{-g}^{(0)}+\delta h_{-g}.
\end{equation}
Here, $h_{g}^{(0)}$ and $h_{-g}^{(0)}$ are large compared to $\delta h_{g}$ and $\delta h_{-g}$, respectively. In addition, we separate $\epsilon = \epsilon^{(0)} +\delta \epsilon$. For the effective model for the Dirac point at $\epsilon =g$, we have $h_{g}^{(0)} = g\mathds{1}_3$, $h_{-g}^{(0)}= -g \mathds{1}_3$, and $\epsilon^{(0)}=g$. The corrections $\delta h_{g}$, $\delta h_{-g}$, and $\delta \epsilon$ are determined by deviations from the band-crossing point, e.g., $\delta \epsilon \sim \hbar v_F k$.

By using the eigenvalue equation $H\Psi = \epsilon \Psi$ with $H$ given in Eq.~(\ref{App-Model-Heff-def}) and $\Psi = \left\{\psi_{g}, \psi_{-g}\right\}$, we can re-express the states $\psi_{-g}$ via the states $\psi_{g}$:
\begin{equation}
\label{App-Model-Heff-psiH}
\psi_{-g} = \left(\epsilon \mathds{1}_3 -h_{-g}\right)^{-1}u^{\dag} \psi_{g}.
\end{equation}
This allows us to write an equation for $\psi_{g}$ only,
\begin{equation}
\label{App-Model-Heff-psiL}
\left[h_{g} +u\left(\epsilon \mathds{1}_3 -h_{-g}\right)^{-1}u^{\dag}\right]\psi_{g} =\epsilon \psi_{g}.
\end{equation}
By using Eq.~(\ref{App-Model-Heff-hL-separate}) and expanding up to the leading nontrivial order in deviations from the band-crossing point at $\epsilon=g$, we obtain
\begin{widetext}
\begin{equation}
\label{App-Model-Heff-eps-h-inverse}
\left(\epsilon \mathds{1}_3 -h_{-g}\right)^{-1} = \left(\epsilon^{(0)} \mathds{1}_3 -h_{-g}^{(0)} +\delta\epsilon \mathds{1}_3 -\delta h_{-g}\right)^{-1} \approx \left[1 -\left(\epsilon^{(0)} \mathds{1}_3 -h_{-g}^{(0)}\right)^{-1} \left(\delta\epsilon \mathds{1}_3 -\delta h_{-g}\right)\right] \left(\epsilon^{(0)} \mathds{1}_3 -h_{-g}^{(0)}\right)^{-1}.
\end{equation}
\end{widetext}
This allows us to rewrite Eq.~(\ref{App-Model-Heff-psiL}) as
\begin{widetext}
\begin{eqnarray}
\label{App-Model-Heff-psiL-2}
&&\left\{h_{g}^{(0)} -\epsilon^{(0)} \mathds{1}_3 +\delta h_{g} +u\left[1 +\left(\epsilon^{(0)} \mathds{1}_3 -h_{-g}^{(0)}\right)^{-1} \delta h_{-g}\right] \left(\epsilon^{(0)} \mathds{1}_3 -h_{-g}^{(0)}\right)^{-1} u^{\dag}\right\}\psi_{g} \nonumber\\
&&=\delta\epsilon \left[\mathds{1}_3 +u\left(\epsilon^{(0)} \mathds{1}_3 -h_{-g}^{(0)}\right)^{-2} u^{\dag}\right] \psi_{g}.
\end{eqnarray}
\end{widetext}
By introducing the wave function $\chi =S^{1/2} \psi_{g}$, which has a proper norm, i.e., $\chi^{\dag}\chi = \psi_{g}^{\dag}\psi_{g} +\psi_{-g}^{\dag}\psi_{-g}$, we rewrite Eq.~(\ref{App-Model-Heff-psiL-2}) in the conventional form $H_{\rm eff}\chi = \delta \epsilon \,\chi$.
Therefore, the effective Hamiltonian describing the states in the vicinity of the threefold band-crossing point $\epsilon=g$ reads
\begin{widetext}
\begin{equation}
\label{App-Model-Heff-Heff-fin-1}
H_{\rm eff} = S^{-1/2} \left\{h_{g}^{(0)} -\epsilon^{(0)} \mathds{1}_3 +\delta h_{g} +u\left[\mathds{1}_3 +\left(\epsilon^{(0)} \mathds{1}_3 -h_{-g}^{(0)}\right)^{-1} \delta h_{-g}\right] \left(\epsilon^{(0)} \mathds{1}_3 -h_{-g}^{(0)}\right)^{-1} u^{\dag}\right\} S^{-1/2},
\end{equation}
\end{widetext}
where
\begin{equation}
\label{App-Model-Heff-S-fin-1}
S = \mathds{1}_3 +u\left(\epsilon^{(0)} \mathds{1}_3 -h_{-g}^{(0)}\right)^{-2} u^{\dag}.
\end{equation}
We use Eqs.~(\ref{App-Model-Heff-Heff-fin-1}) and (\ref{App-Model-Heff-S-fin-1}) to derive the effective models in Sec.~\ref{sec:Model-effective}. While the calculations are straightforward, the intermediate expressions are bulky. Therefore, we do not present them here.

\section{Low energy spectral functions and results for \texorpdfstring{$g/t>1$}{g/t>1}}
\label{sec:App-spectral-low}

\begin{figure*}[t]
\centering
\subfigure[\empty]{\includegraphics[width=0.95\textwidth]{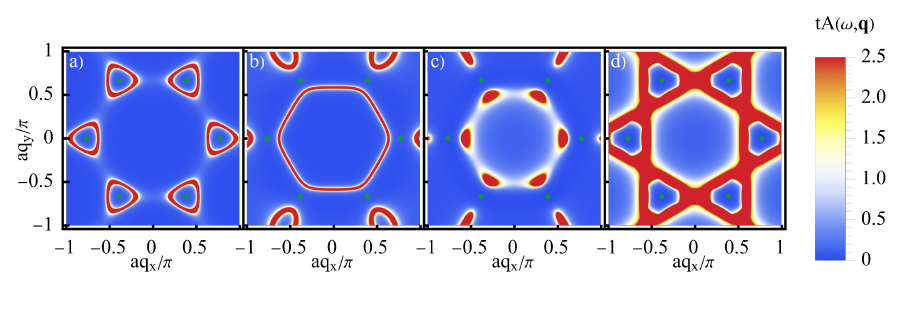}}
\caption{
The spectral functions at $\hbar \omega=0$. We used the aligned $AA-BB-CC$ (panel (a)), hub-aligned $AB-BA-CC$ (panel (b)), mixed $AA-BC-CB$ (panel (c)), and cyclic $AB-BC-CA$ (panel (d)) stackings.
In all panels, we set $g=t$. Green points represent the positions of the band-crossing points. We use tight-binding models with the spectral function defined in Eq.~(\ref{DOS-A-def}) and introduce the phenomenological broadening $\Gamma = 0.05\,t$ by replacing $i0\to i\Gamma$ in the Green function.
}
\label{fig:DOS-Spectral-0}
\end{figure*}

For the sake of completeness, let us also show the spectral function at low energies $\hbar \omega=0$ in Fig.~\ref{fig:DOS-Spectral-0}. As one can see, the low-energy ($\epsilon=0$) spectrum demonstrates nodal rings either surrounding the $K$ points (aligned stacking) or the $\Gamma$ point (hub-aligned stacking); see Figs.~\ref{fig:DOS-Spectral-0}(a) and \ref{fig:DOS-Spectral-0}(b). The mixed stacking is characterized by separated patches. The most intricate, kagome pattern occurs for the cyclic stacking shown in Fig.~\ref{fig:DOS-Spectral-0}(d).

It is instructive also to discuss the case of strong interlayer coupling $g/t\gtrsim1$. It corresponds to a somewhat exotic system where the interlayer coupling constant $g$ is larger than the in-layer hopping parameter $t$. Nevertheless, it might be relevant for artificial systems.

\begin{figure*}[t]
\centering
\subfigure[\empty]{\includegraphics[width=0.99\textwidth]{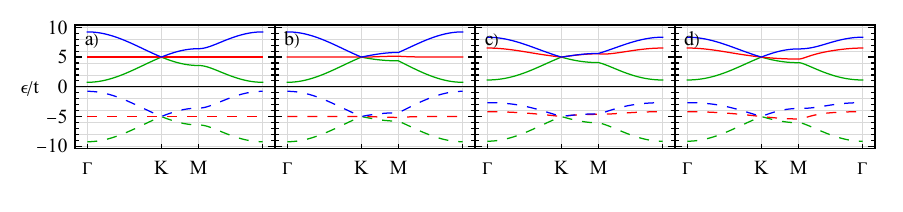}}
\hspace{0.01\textwidth}
\subfigure[\empty]{\includegraphics[width=0.99\textwidth]{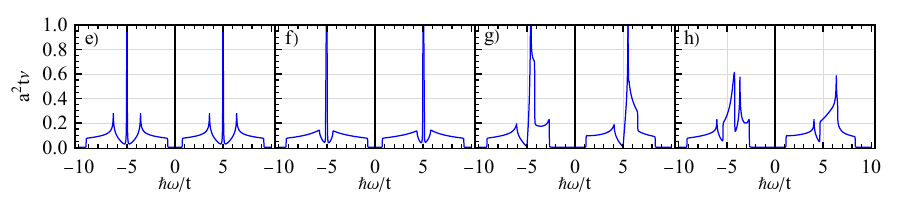}}
\caption{
The energy spectrum (top row) and the corresponding DOS (bottom row) for the tight-binding Hamiltonian (\ref{Model-H-total}) along the $\Gamma-\mbox{K}-\mbox{M}-\Gamma$ line in the Brillouin zone at $g/t=5$. The columns represent the results for the aligned $AA-BB-CC$ (panels (a) and (e)), hub-aligned $AB-BA-CC$ (panels (b) and (f)), mixed $AA-BC-CB$ (panels (c) and (g)), and cyclic $AB-BC-CA$ (panels (d) and (h)) stackings.
}
\label{fig:App-g-large-All}
\end{figure*}

We show the energy spectrum and the DOS for the four nonequivalent stackings in Fig.~\ref{fig:App-g-large-All}. Compared to the case of the smaller coupling constant, cf. Figs.~\ref{fig:Model-effective-aligned-TB-energy}--\ref{fig:Model-effective-cyclic-TB-energy} and Fig.~\ref{fig:DOS-DOS-all}, the spectra and the DOS corresponding to the threefold crossing points at $\epsilon=\pm g$ do not overlap. The shape of the energy spectrum away from the band-crossing points becomes less relevant at larger $g/t$ for the aligned $AA-BB-CC$ stacking. Further, the symmetry of the energy spectrum with respect to the band crossing points becomes evident for the aligned $AA-BB-CC$ and hub-aligned $AB-BA-CC$ stackings; see Figs.~\ref{fig:App-g-large-All}(a), \ref{fig:App-g-large-All}(e), \ref{fig:App-g-large-All}(b), and \ref{fig:App-g-large-All}(f). In agreement with the effective model, the anisotropy of the additional band is suppressed at larger $g$; cf. red lines in Figs.~\ref{fig:Model-effective-hub-TB-energy}(a) and \ref{fig:App-g-large-All}(b). The particle-hole asymmetry and complicated structure of the DOS remain for the mixed $AA-BC-CB$ and cyclic $AB-BC-CA$ stackings; see Figs.~\ref{fig:App-g-large-All}(c), \ref{fig:App-g-large-All}(g), \ref{fig:App-g-large-All}(d), and \ref{fig:App-g-large-All}(h).

\bibliography{library-short}

\end{document}